%% file: CNS2021_file_sharing.tex
\documentclass[conference]{IEEEtran}
\IEEEoverridecommandlockouts
\usepackage{cite}
\usepackage{amsmath,amssymb,amsfonts}
\usepackage{algorithmic}
\usepackage{graphicx}
\usepackage{textcomp}
\usepackage{enumerate}
\usepackage{makecell}
\usepackage{mathrsfs}
\usepackage{tabularx}
\usepackage{framed}
\usepackage{tcolorbox}
\usepackage{xcolor}
\def\BibTeX{{\rm B\kern-.05em{\sc i\kern-.025em b}\kern-.08em
    T\kern-.1667em\lower.7ex\hbox{E}\kern-.125emX}}

\makeatletter
\def\ps@headings{%
\def\@oddhead{\mbox{}\scriptsize\rightmark \hfil \thepage}%
\def\@evenhead{\scriptsize\thepage \hfil \leftmark\mbox{}}%
\def\@oddfoot{}%
\def\@evenfoot{}}
\makeatother
\def\BibTeX{{\rm B\kern-.05em{\sc i\kern-.025em b}\kern-.08em
    T\kern-.1667em\lower.7ex\hbox{E}\kern-.125emX}}

\usepackage{url}

\def\BibTeX{{\rm B\kern-.05em{\sc i\kern-.025em b}\kern-.08em
    T\kern-.1667em\lower.7ex\hbox{E}\kern-.125emX}}

\newcommand{\delete}[1]{}

\makeatletter 
  \newcommand\figcaption{\def\@captype{figure}\caption} 
  \newcommand\tabcaption{\def\@captype{table}\caption} 
\makeatother

\begin{document}

\title{BBS: A Blockchain Big-Data Sharing System}

\author{
\IEEEauthorblockN{Shan Wang\IEEEauthorrefmark{1}\IEEEauthorrefmark{3},
Ming Yang\IEEEauthorrefmark{1},
Tingjian Ge\IEEEauthorrefmark{3},
Yan Luo\IEEEauthorrefmark{3},
Xinwen Fu\IEEEauthorrefmark{3}
}
\IEEEauthorblockA{\IEEEauthorrefmark{1}
Southeast University. Email: shanwangsec@gmail.com, yangming2002@seu.edu.cn}
\IEEEauthorblockA{\IEEEauthorrefmark{3}University of Massachusetts Lowell. Email:\{Yan\_Luo, Xinwen\_Fu\}@uml.edu, ge@cs.uml.edu}
}

\maketitle

\begin{abstract}
Chain of custody is needed to document the sequence of custody of sensitive big data.
In this paper, we design a blockchain big-data sharing system (BBS) based on Hyperledger Fabric. 
We denote the data stored outside of a ledger for sharing as ``off-state" and ``big data'' (referring to extremely large data) is in this category.
In our off-state sharing protocol, a sender registers a file with BBS for sharing. To acquire the file, an authenticated and authorized receiver has to use transactions and interacts with BBS in four phases, including the file transfer request,
encrypted file transfer, key retrieval, and file decryption.
The corresponding transactions are recorded in the ledger and serve as chain of custody to document the trail of the data.
Compared with related work, BBS can perform the four phases autonomously.
It utilizes the permissioned blockchain, i.e. Hyperledger Fabric, for access control and can defeat dishonest receivers.
We design and implement a prototype of BBS for big file sharing. Extensive experiments were performed to validate its feasibility and performance. 
\end{abstract}


\input{sections/sec1_introduction}

\input{sections/sec2_background}

\input{sections/sec3_model}

\input{sections/sec4_method}

\input{sections/sec6_evaluation}
\input{sections/sec9_conclusion}




\bibliographystyle{IEEEtran}
\bibliography{file_sharing.bib}

\end{document}

%% file: sections/sec1_introduction.tex
\section{Introduction}

A blockchain system can build trust in the data that it maintains without a centralized authority. Data in conventional blockchain systems is often stored in a ledger, which includes a world state and a blockchain. The world state stores the current system state such as the user cryptocurrency balance in Bitcoin \cite{antonopoulos2014mastering} and the blockchain saves all transaction history, which contains operations on the world state and/or the data used to update the world state. Smart contract controls operations on the world state. The whole transaction history produces the current state values and can work as auditing evidence. The ledger is synchronized across all blockchain nodes.

In this paper, we use blockchain to share sensitive big data such as scientific and biomedical data freely and establish the chain of custody \cite{Custody_Wikipedia}. Sharing such data with trusted parties without charge allows independent verification of published scientific results and enhances opportunities for new discoveries. With concerns of intellectual property (IP) theft and industrial espionage \cite{budd::data-theft::2019}, we desire secure and trustworthy big data sharing systems that can record the chain of custody to document the trail of the data, e.g. who requests and owns what data. \looseness=-1



However, existing blockchain frameworks cannot be directly applied to big data sharing. In current blockchain frameworks, the data size and data type in ledgers is limited due to transaction fees, system performance and other concerns \cite{BitcoinWeakness,BitcoinForum,GasLimit,gorenflo2020fastfabric}.
The ledger is conventionally designed to maintain the state data, such as cryptocurrency balance.
All blockchain nodes often maintain the same ledger. However, because of privacy and intellectual property (IP) concerns, owners may not want to share the big data across all nodes. 


Related work on big data sharing pertaining to Blockchain cannot be used for the application we target. FairSwap \cite{dziembowski2018fairswap} is an off-chain based big file selling application. 
(\romannumeral 1) FairSwap sells digital goods for money in Ethereum with cryptocurrency. 
Users need pay transaction fees to miners for smart contract execution. It is not designed for free data sharing for scientific discovery. 
(\romannumeral 2) In FairSwap, the file transfer and encryption/decryption operations are conducted off-chain by a sender and a receiver while the blockchain system conducts work such as cryptocurrency transfer and encryption key exchange. That is, FairSwap segregates file transfer from the blockchain system and is not designed for autonomous big data sharing.
(\romannumeral 3) FairSwap works on a public blockchain. The encryption key is revealed to the public on the ledger. Once an encrypted file is leaked, for example, intercepted by malicious cyber players, the encrypted file can then be decrypted. 
A token-based off-chain data sharing scheme is briefly discussed in \cite{zhang2018fhirchain} and there is no concrete protocol. Their blockchain system has no control over the actual data sharing process. Chain of custody cannot be properly established.


We propose a blockchain big-data sharing system (BBS) based on the permissioned blockchain framework, the Hyperledger Fabric~\cite{androulaki2018hyperledger} (abbreviated as {\em Fabric} in the rest of the paper), so as to 
securely share sensitive big data with authenticated and authorized users and record the chain of custody within the ledger. Our major contributions can be summarized as follows.
We introduce the concept ``off-state", which is data, particularly big data, maintained at a separate storage space from the ledger at blockchain nodes. 
Off-states can be shared between parties of interest and do not need to be synchronized across all nodes. Smart contracts can directly operate on the off-states such as sharing. \looseness=-1

To autonomously and securely establish the chain of custody of off-states, we propose a novel off-state sharing protocol utilizing the smart contract.
Only an authorized receiver can obtain off-states through transactions, which document the chain of custody.
Requested data is encrypted and transferred on demand. The receiver has to propose a transaction, which will be recorded into the ledger, to obtain the key and decrypt the file. The key is known only to the sender and receiver.
Therefore, dishonest receivers who obtain the key cannot deny that they obtain the original data.

We implement a prototypical BBS with Fabric running our off-state data sharing protocol. Extensive experiments are performed to evaluate BBS' feasibility and performance such as latency of the big file sharing between pairs of blockchain nodes.
BBS performs similarly or better compared to SFTP/SCP applications in Linux. A long transaction involving big file transfer does not block other operations of BBS, which can process multiple big file transfer and other transactions simultaneously. 
\looseness=-1

The rest of this paper is organized as follows. Section \ref{sec::background} introduces the background knowledge---Hyperledger Fabric. Section \ref{sec::system Model} presents the challenges, system model and threat model for off-state sharing. The off-state sharing protocol is introduced in Section \ref{sec::Integration_Scheme}. We evaluate the prototypical BBS in Section \ref{sec::Evaluations}
and conclude this paper in Section \ref{sec::Conclusion}.

%% file: sections/sec2_background.tex
\section{Background}
\label{sec::background}

In this section, we introduce the Hyperledger Fabric, a permissioned blockchain framework. 


\subsection{Transaction Workflow}
\label{subsubsec::FabricTransactionWorkflow}

Fabric has three types of nodes, i.e. peers, orderers and clients in charge of different tasks during the transaction workflow. Peers maintain ledgers and smart contracts, and perform as the backbone of a Fabric network. Involved parties need to approve the deployment of a smart contract. 

Fabric adopts a three-phase ``execute-order-validate'' transaction workflow as shown in Fig. \ref{fig:txworkflow}: (\romannumeral 1) A client/user proposes a transaction {\em proposal} to the endorsers, which are a subset of peers and specified by an endorsement policy; (\romannumeral 2) Every endorser independently executes the chaincode, signs the execution results, and returns the results with a corresponding signature to the client as a proposal response. Note that the execution results are not updated to the world state at this stage; (\romannumeral 3) If the returned execution results from different endorsers are the same, {\em the client} constructs a transaction, which contains both the transaction proposal, execution results and a list of signatures (endorsements). 
The client sends the transaction to the orderer nodes. (\romannumeral 4) The orderer nodes bundle collected transactions into a new block, and distribute the new block to all peers including both endorsers and non-endorsers. (\romannumeral 5) All peers validate the transactions in the received new block. For one transaction, if its endorsements pass the endorsement policy check and its execution results pass the version conflict check, each peer will update the world state according to execution results. 
After validating all transactions, each peer appends the new block to its local blockchain.


\subsection{Access Control Mechanism}

{\bf Membership Service}:
Fabric issues each user or node a certificate as the identity, which is bonded with one organization within the consortium of organizations who build the Fabric system. A node or a user represents an organization. Only authorized nodes and users can participate in a Fabric system.\looseness=-1

{\bf Fine-grained Data Isolation}:
Fabric introduces a fine-grained data isolation mechanism, i.e., private data collection (PDC). PDC data is sensitive and shared among only a subset of peer nodes, i.e. a subset of organizations. The PDC is stored in the world state. Only PDC member peers store the original data, while PDC non-member peers only store the data hashes.


{\bf Multi-level Endorsement Policy}:
The endorsement policy setting is flexible in Fabric.
An endorsement policy stipulates which peers need to perform as endorsers to endorse a transaction. Endorsing involves executing smart contracts and signing the execution results.
Each smart contract has a default chaincode-level endorsement policy that manages all public data and PDC data in the world state. A key-level endorsement policy can be customized to manage one specific key-value data. A collection-level endorsement policy can be used to manage the PDC data.

%% file: sections/sec3_model.tex
\section{Off-State Data Sharing Models}
\label{sec::system Model}


Our goal is to design a blockchain based big data sharing system that can establish the chain of custody of the big data, being able to track the trail. The ledger provides non-repudiation evidences.
The big data is shared between a sender and a receiver within a consortium of organizations. In this section, we first present the challenges and then present our system model and threat model.


\subsection{Challenges}
\label{sec::challenge}


We identify three challenges for designing BBS.

{\em Challenge 1---Storage Space Limitation}:
Due to the storage limitations of ledgers, it is impractical to store big data in ledgers or pack big data in transactions. 

{\em Challenge 2---Privacy Requirement}:
Big data may be sensitive. The owner may not want to share the data with everyone. Access control shall be adopted.
However, in a conventional blockchain system, all nodes maintain the same data. Therefore, the blockchain system cannot be directly used for big data sharing.

{\em Challenge 3---Security Requirement}: 
In a big-data sharing session, the sender or the receiver may misbehave. 
The receiver may dishonestly deny that she/he has received the data from the sender.
The sender may share the data not consistent with the description that she/he provides.
If a dispute occurs, the blockchain shall provide evidences for the case.

\begin{figure*}[!ht]
\begin{minipage}[c]{0.68\columnwidth}
\centering
\includegraphics[height=0.75\columnwidth]{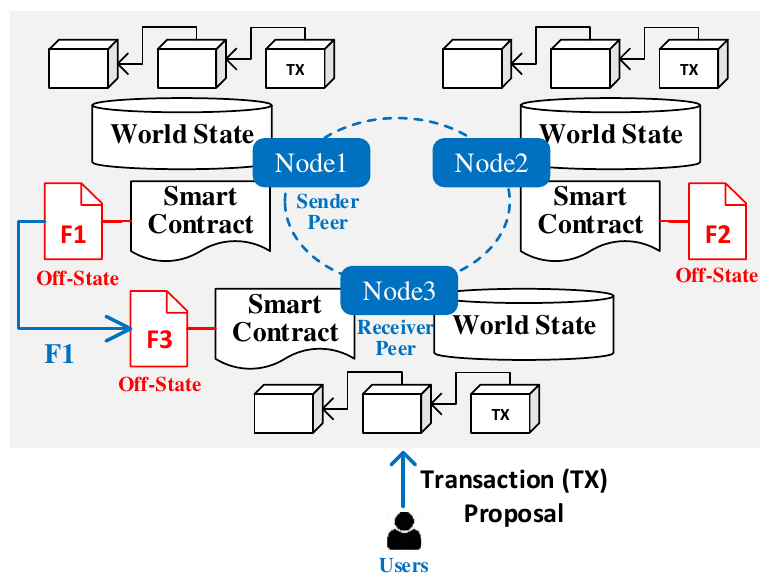}
\caption{Blockchain-based off-state sharing model}
\label{fig:system_model} 
\vspace{-2mm}
\end{minipage}
\hspace{0.1cm}
\begin{minipage}[c]{0.68\columnwidth}
\centering
\includegraphics[height=0.75\columnwidth]{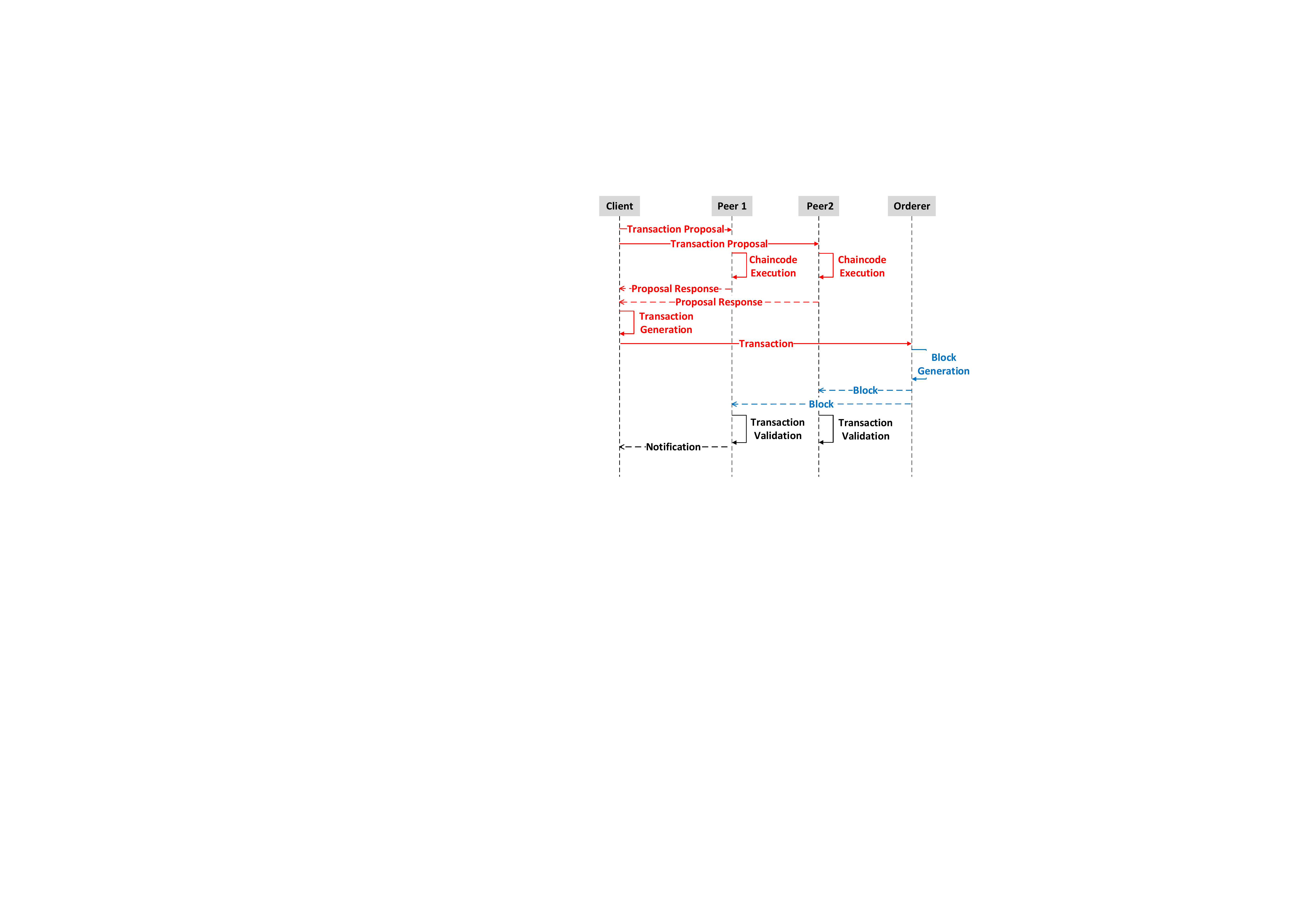}
\caption{Three-phase transaction workflow in Fabric}
\label{fig:txworkflow} 
\vspace{-2mm}
\end{minipage}
\hspace{0.1cm}
\begin{minipage}[c]{0.68\columnwidth}
\centering
\includegraphics[height=0.75\columnwidth]{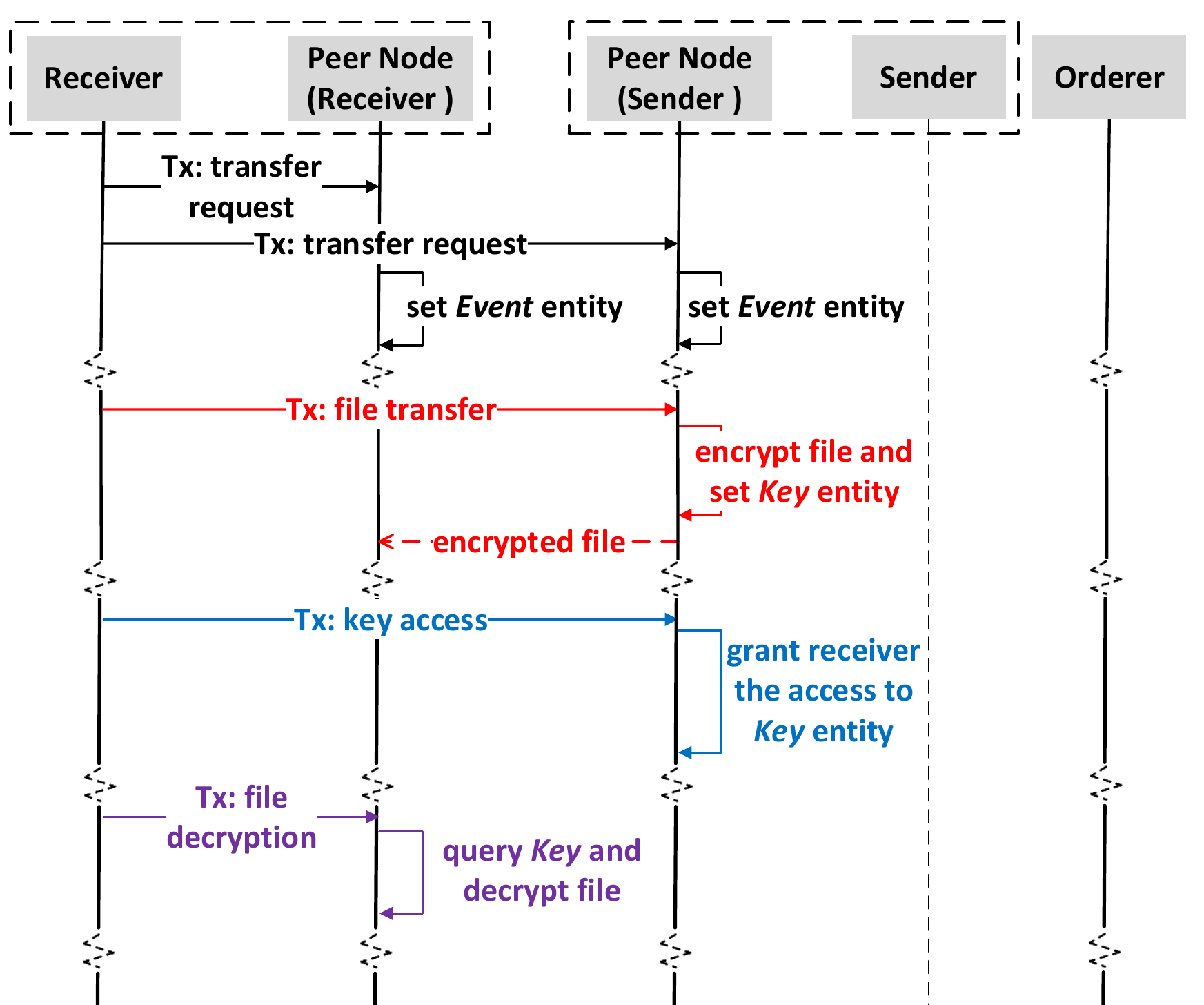}
\caption{Four phases in off-state sharing protocol}
\label{fig::Module2} 
\vspace{-2mm}
\end{minipage}
\end{figure*} 

\subsection{System Model}
\label{sec::off-stateConcept}


\textit{Permissioned Blockchain Settings:}
A consortium of organizations $\mathcal{O}=\{O_1,...,O_n\}$ cooperate together to build a permissioned blockchain system $\mathcal{B}$. In Fabric, an organization can have multiple users. For clarity and simplicity but without loss of generality, assume each organization contributes one peer node, and authorizes one user to join the system.
The peer node set is denoted as $\mathcal{N}=\{N_1,...,N_n\}$. The user set is denoted as $\mathcal{U}=\{U_1,...,U_n\}$. A user interacts with $\mathcal{B}$ through a client node. $N_i$ and $U_i$ belong to $O_i$.
The orderer node setting will be discussed in Section \ref{sec::systemFramework}.
All organizations cooperate to develop and deploy the smart contract $C$ on peers $\mathcal{N}$ to implement a common business goal such as big data sharing. 
 
We introduce the concept of ``off-state", which is stored at blockchain nodes $\mathcal{N}$, but outside of the ledgers, particularly the world state. For example, the off-state can be big data. Therefore, we generalize big data sharing as off-state data sharing. 
Off-states can be inconsistent across $\{N_1,...,N_n\}$, and shared between pairs of nodes.
A smart contract $C$ can directly operate on the off-state for read, write, encryption/decryption and other operations including data sharing.

Fig. \ref{fig:system_model} shows our blockchain off-state sharing system model.
Each node maintains a world state database and an off-state storage space when needed.
Users propose transactions to trigger the smart contracts to transfer the off-state data. An off-state sharing session occurs between two organizations, $O^S$ and $O^R$, within $\mathcal{O}$. We denote their nodes as $N^S$ and $N^R$, and their users as $S$ and $R$ respectively.
The smart contract fully controls the sharing process and can transfer off-state data $f$ from $N^S$ to $N^R$ through an off-state sharing protocol.


\subsection{Threat Model}
\label{sec::securityModel}
We assume that the underlying blockchain infrastructure $\mathcal{B}$ including all peers and orderers is secure. A user $U_i$ only can retrieve files in his/her off-state storage space of $N_i$.
In Fabric, the smart contract is deployed in a decentralized way.
No organization can individually control the smart contract. Therefore, we assume the smart contract $C$ is trusted.

Users $S$ and $R$ may be dishonest. A user interacts with $\mathcal{B}$ through a client node. A user controls the application that runs on the client node, and may maliciously change the application. For example, receiver $R$ sends a transaction proposal and requests $C$ to transfer $f$ from sender $S$ to $R$. according to the transaction workflow in Section \ref{subsubsec::FabricTransactionWorkflow}, the smart contract executes before the transaction ordering and validation phases. After the smart contract transfers the data $f$, a dishonest $R$ may change the client application, and does not submit the transaction for ordering. Then the blockchain cannot record this data transfer activity but $R$ has received $f$.

%% file: sections/sec4_method.tex
\section{Off-State Sharing Protocol}
\label{sec::Integration_Scheme}

In this section, we first present our off-state sharing protocol based on Fabric, and then discuss how it addresses the three challenges in Section \ref{sec::challenge}. 

\subsection{Overview}
The off-state sharing protocol has two stages. We use a big file $f$ as an off-state data example. 
In Stage 1, sender $S$ prepares the big data $f$ to be shared. $f$ is stored in the off-state of sender node $N^S$.
In Stage 2, blockchain system $\mathcal{B}$ takes over the whole off-state sharing process, and autonomously and securely establish the chain of custody. Receiver $R$ only interacts with $\mathcal{B}$ through transactions, and does not need to wait for response from $S$. Smart contract (i.e., chaincode called by Fabric) $C$ encrypts $f$, transfers the encrypted data $f_z$ from the off-state of $N^S$ to the off-state of $N^R$, and stores the symmetric encryption key $k$ in the world state. $k$ is managed using a PDC of Fabric. $R$ has to request $k$ through a transaction in order to decrypt $f_z$. If $R$ is legitimate, 
$C$ will reveal $k$ to $R$. 
$C$ puts critical information in the execution results, which will be used to update the world state and will be recorded in transactions which perform as evidences.

\begin{figure}
\centering
\includegraphics[width=1\columnwidth]{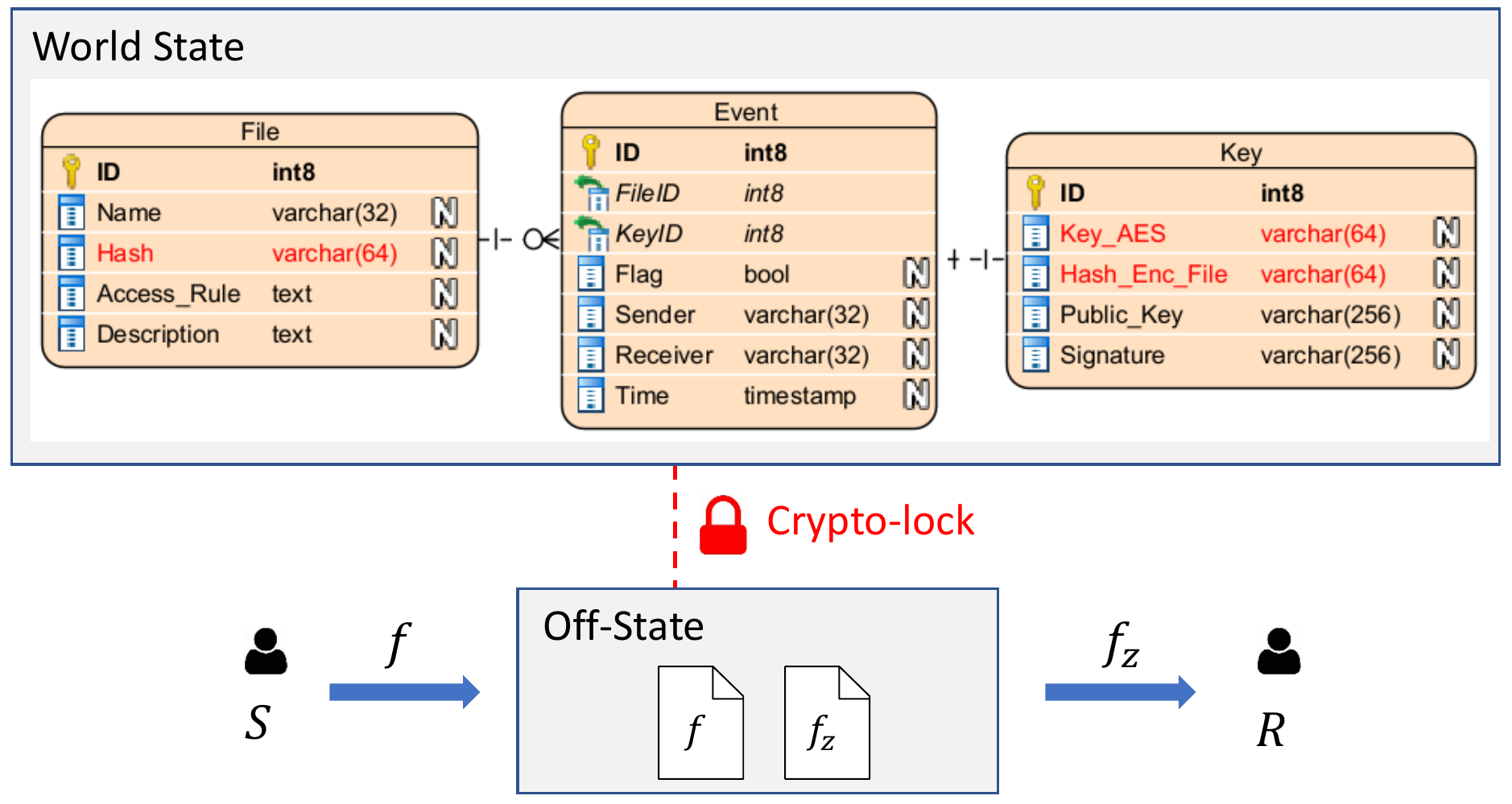}
\caption{World state and off-state design in the off-state sharing protocol. The off-state is bound with the world state through hashes and encryption keys which are called cryptographic locks
}
\label{fig::Module2DataPlane} 
\end{figure}

\subsection{Stage 1: Preparing Off-State Data}
An sender and owner $S$ prepares the big file to be shared. $S$ uploads $f$ to the off-state storage of $N^S$. $S$ also proposes a transaction---``Tx: File upload"---to trigger smart contract $C$ to calculate and record the metadata of $f$ into the world state.
Fig. \ref{fig::Module2DataPlane} illustrates the \texttt{File} entity, i.e. \texttt{ $\langle$ID, Name, Hash, Access\_Rule, Description$\rangle$}, in the world state.

\begin{itemize}
\item The file hash $h$, calculated and maintained by the blockchain, can be used to verify the integrity of $f$. $C$ calculates and records $h$ in the world state. Every node maintains $h$.

\item The access rules protect the privacy of $f$. $S$ individually defines the access rules based on the identities in the permissioned blockchain. For example, a rule $\{O_1, O_2\}$ means $f$ only can be shared with organization $1$ and organization $2$. 

\item The file entity in the world state database is public to all users for searching. 
\end{itemize}
Once the \texttt{File} entity is set up, $\mathcal{B}$ takes over the file sharing process. $\mathcal{B}$ autonomously shares files following the access rule defined by $S$.

\subsection{Stage 2: Sharing Off-State Data}
\label{Sharing_Off_State_Data}

Fig. \ref{fig::Module2} shows the workflow of sharing a big file over Fabric. 
The sharing process involves four transactions. These four transactions follow the same transaction workflow as introduced in Section \ref{subsubsec::FabricTransactionWorkflow}. However endorsers for each transaction are different according to the specific endorsement policy setting. An endorsement policy stipulates who should sign the transaction. Fig. \ref{fig::Module2} shows only transaction steps of interest while ignoring other steps such as ordering of the transaction workflow in Fig. \ref{fig:txworkflow}.


\subsubsection{Phase 1: File Transfer Request}
$R$ shall request the transfer of a particular file $f$ stored at $N^S$ and the request shall pass the access control governed by a specific endorsement policy.
$C$ can get the identity of the user who sends the transaction proposal. $C$ running at an endorser peer checks if the user identity satisfies the access rule in the \texttt{File} entity, and records the results in an \texttt{Event} entity in the world state. 
An \texttt{Event} entity is in the form of \texttt{ $\langle$ID, FileID, KeyID, Flag, Sender, Receiver, Time$\rangle$} as shown in Fig. \ref{fig::Module2DataPlane}, where the $Flag$ is a Boolean value indicating if the endorser permits the file transfer. 
The endorsement policy can be set as $AND(O^S,O^R)$, which stipulates peers from both $O^S$ and $O^R$ should endorse the transaction.
A valid transaction means that $O^S$ and $O^R$ reach a consensus and sign on the same results (i.e. \texttt{Event} entry), either denial of or agreement on the file transfer. We call the corresponding transaction ``Tx: transfer request".\looseness=-1


\subsubsection{Phase 2: Encrypted File Transfer}
After the file transfer request is approved, $R$ proposes another transaction---``Tx: file transfer"---to initiate the file transfer. The endorsement policy for ``Tx: file transfer" can be set as $AND(O^S)$,
which means the sender peer $N^S$ works as an endorser and signs the transaction.
$C$ in $N^S$ first checks $Flag$ in the \texttt{Event} entity. The file transfer continues only if $Flag$ is $True$.

$C$ in $N^S$ then encrypts $f$ using $k$, signs the encrypted file $f_z$ using an asymmetric key $(pk, sk)$, and sends $f_z$ to the off-state of $N^R$ via a file transfer protocol such as \textit{SFTP}. The encryption key and the auxiliary information of $f_z$ are put in the world state as an \texttt{Key} entity. The \texttt{Key} entity is in the form of \texttt{$\langle$ID, Key\_AES, Hash\_Enc\_File, Public\_Key, Signature$\rangle$}. 
An instance of \texttt{Key} entity is \texttt{$\langle$ $k$, $h_z$, $pk$, $s$ $\rangle$}, where $h_z$ is the hash of $f_z$. $s=Sign(sk,f_z)$, where the function $Sign$ uses $sk$ to sign $f_z$ and outputs the signature $s$.
Specifically in Fabric, $k$ is set as a PDC data. PDC can guarantee that only PDC members can see the actual encryption key and others only know its hash. In this phase, only $O^S$ is the PDC member. That's only $N^S$ keeps the original $k$, and $N^R$ only has the hash of $k$. 

\subsubsection{Phase 3: Key Retrieval}
$R$ first checks if $N^R$ has received $f_z$ and if $f_z$ matches with $h_z$ and $s$. $R$ can send a query request to $N^R$ and delegate $C$ to check $f_z$. In Fabric, the query request will not generate a transaction. Only after $N^R$ has received the file does $R$ propose a transaction---``Tx: key access"---to $\mathcal{B}$ to request $k$.
The endorsement policy for ``Tx: key access" is set as $AND(O^S)$ so that $N^S$ performs as an endorser. $C$ in $N^S$ checks if $R$ is the legitimate receiver of the corresponding \texttt{Event} entity. If $R$ is, $C$ sets $k$ as a new PDC data whose members include both $O^S$ and $O^R$. Then the PDC mechanism will store $k$ in both $N^S$ and $N^R$. Therefore, $N^R$ will be able to retrieve $k$. Note that the corresponding transaction only stores PDC hash. The PDC value is transmitted using a specific protocol in Fabric. We assume the PDC mechanism is reliable and secure since the PDC mechanism is a part of our underlying blockchain framework. That is $k$ is updated to the world state database of $N^R$ only after the transaction is validated as valid in the blockchain. 
PDC also ensures that $k$ is definitely shared with the legitimate $R$ and not publicly disclosed.

\subsubsection{Phase 4: File Decryption}
Finally, $R$ proposes the last transaction ---``Tx: file decryption"---to decrypt $f_z$. In Fabric, the endorsement policy can be set as $AND(O^R)$ so that $N^R$ performs as an endorser. $N^R$ can access $f_z$ in its off-state storage and access $k$ in its world state. $N^R$ uses $k$ to decrypt $f_z$ to get $f$, and further verifies if $f$ matches with $h$ in the \texttt{File} entity.

\subsection{Analysis}
\label{sec::stage3}

Our off-state sharing protocol and system addresses the three challenges that are introduced in Section \ref{sec::challenge} as follows.

{\em Challenge 1---storage space limitation}: We introduce the concept of ``off-state" and store the big data off the ledger. 

{\em Challenge 2---privacy requirement}: Our protocol involves access control mechanisms to allow off-states to be shared between pairs of nodes, not across all blockchain nodes. Besides, an owner can define access rules to control the data sharing permission.
This feature also saves storage space at nodes that do not need the off-state data.

{\em Challenge 3---security requirement}. Our protocol securely establishes the chain of custody of the off-state data. 
(\romannumeral 1) The critical information such as $h$ and $h_z$  generated during the data sharing process is recorded in the corresponding transactions and the world state. $S$ cannot deny the content of $f$ which has been shared unless $S$ finds a collision of the cryptographic hash function. 
(\romannumeral 2) The built-in PDC mechanism can guarantee that $R$ only knows the encrypted file $f_z$ and the hash of $k$ if $R$ does not submit the transaction---``Tx: key access". Only after ``Tx: key access" is validated as valid and recorded into the blockchain, $R$ can get $k$. $R$ has to propose and submit ``Tx: key access" to get $k$ to decrypt $f_z$ unless $R$ cracks the hash of $k$.
The trusted $C$ generates $h$, $h_z$ and $k$, and can guarantee their correctness. 
Therefore, $S$ cannot deny the content of the shared $f$ and $R$ cannot deny she/he has received $f$. The blockchain provides non-repudiation evidences. 




%% file: sections/sec6_evaluation.tex
\section{Evaluation}
\label{sec::Evaluations}

We have implemented a prototypical off-state sharing system, {\em BBS}, based on Fabric, and evaluate its feasibility and performance in this section.

\subsection{Experiment Setup}
\label{sec::systemFramework}

\begin{figure}
\centering
\includegraphics[width=1\columnwidth]{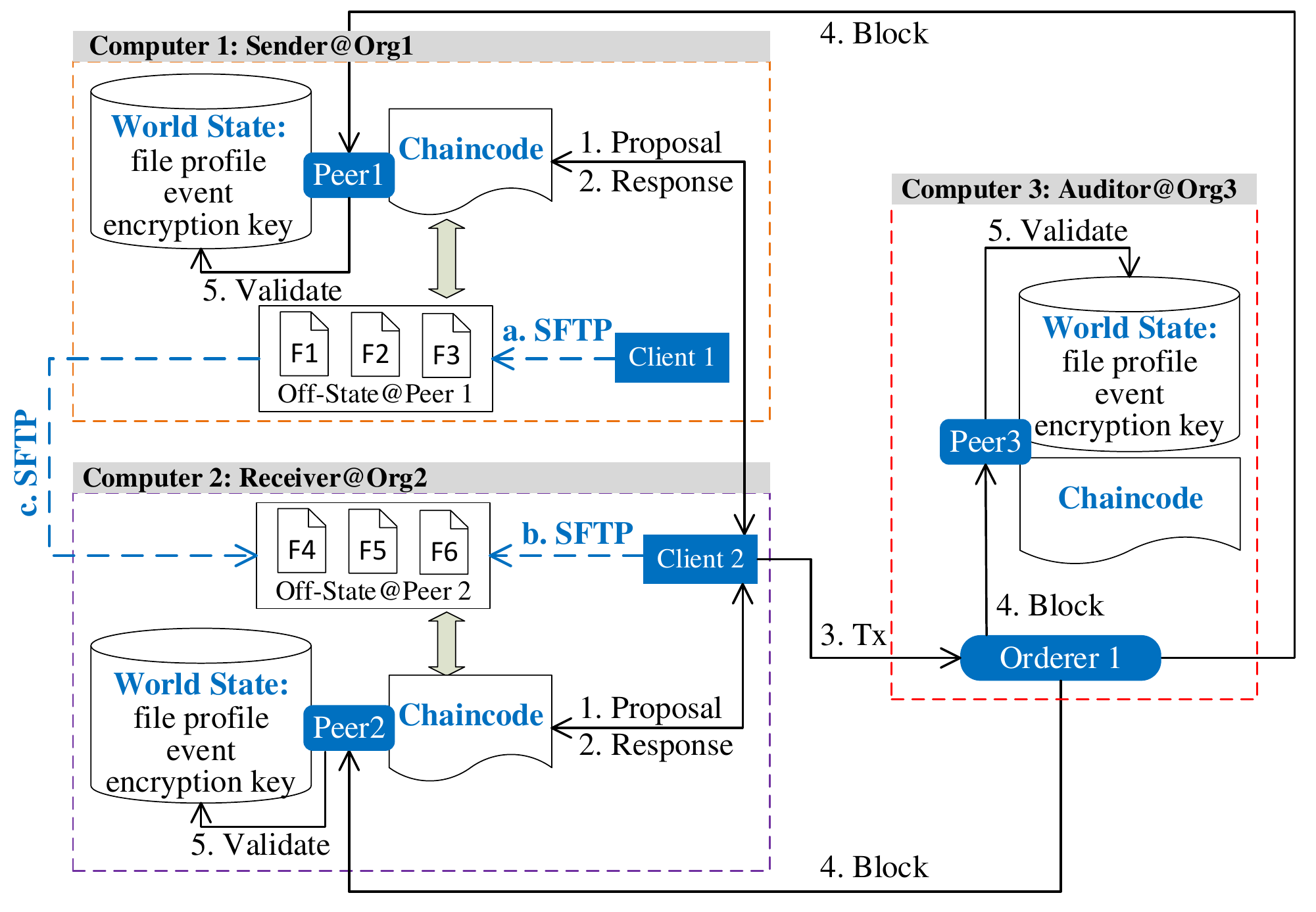}
\caption{Prototypical off-state sharing system based on Hyperledger Fabric
}
\label{fig::framework} 
\end{figure}

Fig. \ref{fig::framework} shows the prototypical BBS, which has three organizations $\mathcal{O}=\{O_1,O_2,O_3\}$. Each party contributes one physical computer. These three computers access the network through WiFi.
$O_1$ and $O_2$ perform as the sender and the receiver. The sender computer and the receiver computer are located in different buildings on a university campus. Fabric nodes run in Docker containers. The sender computer and the receiver computer respectively run one peer node and one client node. The computer of $O_3$ runs one peer node and one orderer node. We set a single orderer node, which performs the equivalent functions as a group of orderer nodes, according to the official Fabric test-network.
We integrate an off-state storage and the \textit{SFTP} \cite{sftp_Wikipedia} service into peer nodes of interest. The smart contract is called chaincode in Fabric. The chaincode is developed using \textit{Golang}. \looseness=-1
The chaincode runs in peer nodes, and can transfer and retrieve files from the local off-state storage.
We use \textit{Node.js} to develop the client application that a user uses to propose transactions to the peer nodes. The node.js application automatically runs one off-state sharing session.

\begin{figure*}[!ht]
\begin{minipage}[c]{0.65\columnwidth}
\centering
\includegraphics[width=1\columnwidth]{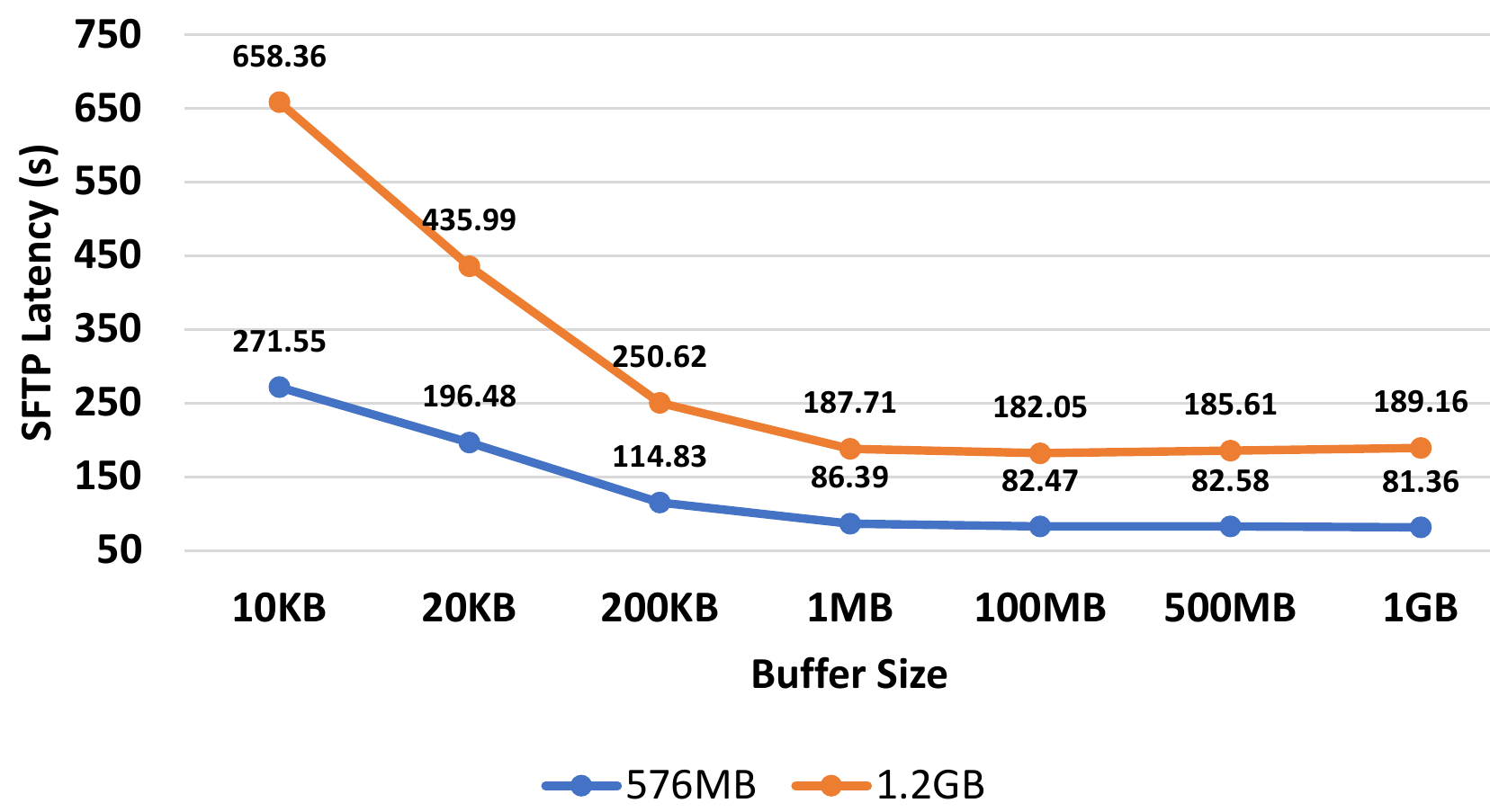}
\caption{\textit{SFTP} latency versus buffer size
}
\label{fig::SFTP_buffer} 
\end{minipage}
\hspace{0.3cm}
\begin{minipage}[c]{0.65\columnwidth}
\tabcaption{Test File List}
\resizebox{\textwidth}{!}{
\begin{tabular}{lll}
\hline
Type & Size  & Describtion                       \\ \hline
.pdf & 67MB  & ``C++ Primer Plus" eBook          \\
.mp4 & 218MB & An over 5 hours song list video   \\
.tif & 576MB & The image of Moon from NASA       \\
.zip & 1.2GB & A collection of medical images    \\
.rar & 2.6GB & The compressed file of one movie  \\
.zip & 5.3GB & The compressed file of two movies \\ \hline
\end{tabular}}
\label{table::file_list}
\end{minipage}
\hspace{0.3cm}
\begin{minipage}[c]{0.65\columnwidth}
\centering
\tabcaption{Percentage of SFTP Latency in ``Tx: file transfer" Latency}
\resizebox{\textwidth}{!}{
\begin{tabular}{c|cccc}
\hline
Size                       & 1 Parallel & 2 Parallel & 4 Parallel & 8 Parallel \\ \hline
\multicolumn{1}{c|}{67MB}  & 75.30\%    & 85.33\%    & 93.38\%    & 92.01\%    \\
\multicolumn{1}{c|}{218MB} & 88.05\%    & 92.82\%    & 95.73\%    & 84.13\%    \\
\multicolumn{1}{c|}{576MB} & 92.43\%    & 94.86\%    & 87.50\%    & 82.33\%    \\
\multicolumn{1}{c|}{1.2GB} & 91.18\%    & 85.13\%    & 83.86\%    & 83.71\%    \\
\multicolumn{1}{c|}{2.6GB} & 86.24\%    & 85.71\%    & 84.36\%    & 84.08\%    \\
\multicolumn{1}{c|}{5.3GB} & 85.84\%    & 83.56\%    & 81.39\%    & 82.68\%    \\ \hline
\end{tabular}}
\label{table::sftpPercentTran}
\end{minipage}
\end{figure*}

\subsection{Performance and Feasibility}

We use \textit{Golang}'s \textit{SFTP} package to perform file transfer and find that the sender's buffer size parameter affects speed and latency of \textit{SFTP}. We set the buffer size as 1MB in our system based on Fig. \ref{fig::SFTP_buffer}, which measures the file transfer latency versus buffer size. A larger buffer size than 1MB does not help the latency much any more.

We use files of different types and various sizes to evaluate BBS' feasibility and performance. Table \ref{table::file_list} lists those test files. Recall that a file sharing session involves four transactions as shown in Fig. \ref{fig::Module2}. We evaluate the latency of different transactions and the latency of the whole session. 

\begin{figure*}[!ht]
\begin{minipage}[c]{0.45\columnwidth}
\centering
\includegraphics[height=0.72\textwidth]{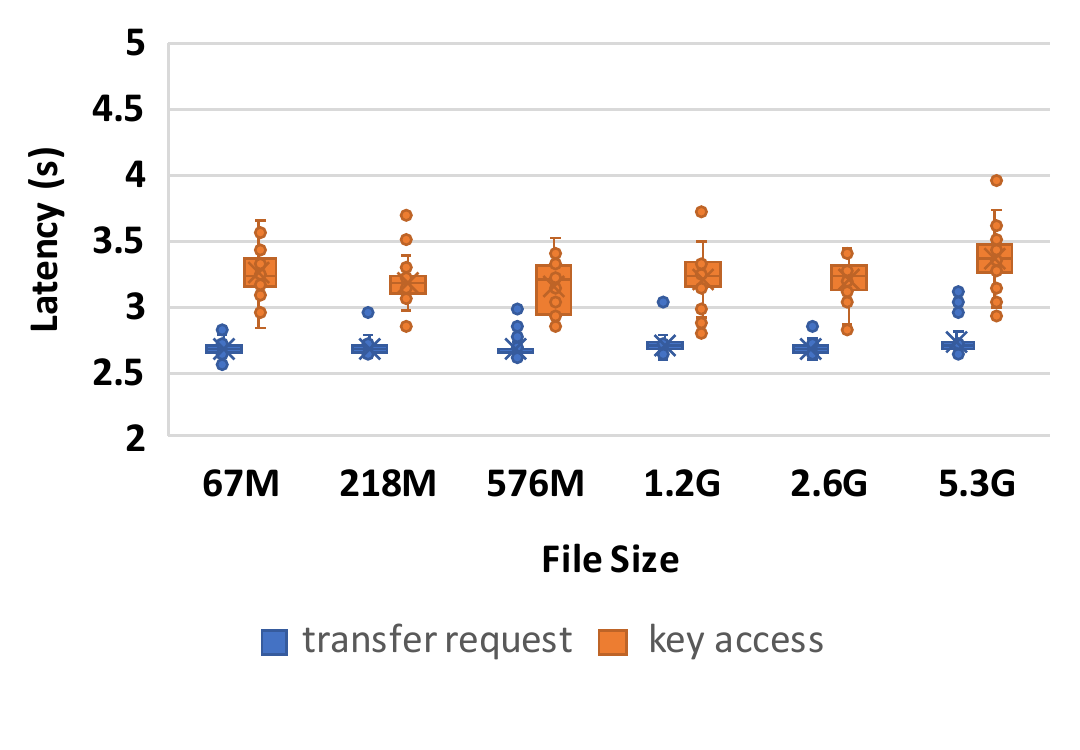}
\caption{Transaction latency of ``Tx: transfer request" and ``Tx: key access"}
\label{fig::1par_req} 
\end{minipage}
\hspace{0.1cm}
\begin{minipage}[c]{0.45\columnwidth}
\centering
\includegraphics[height=0.72\textwidth]{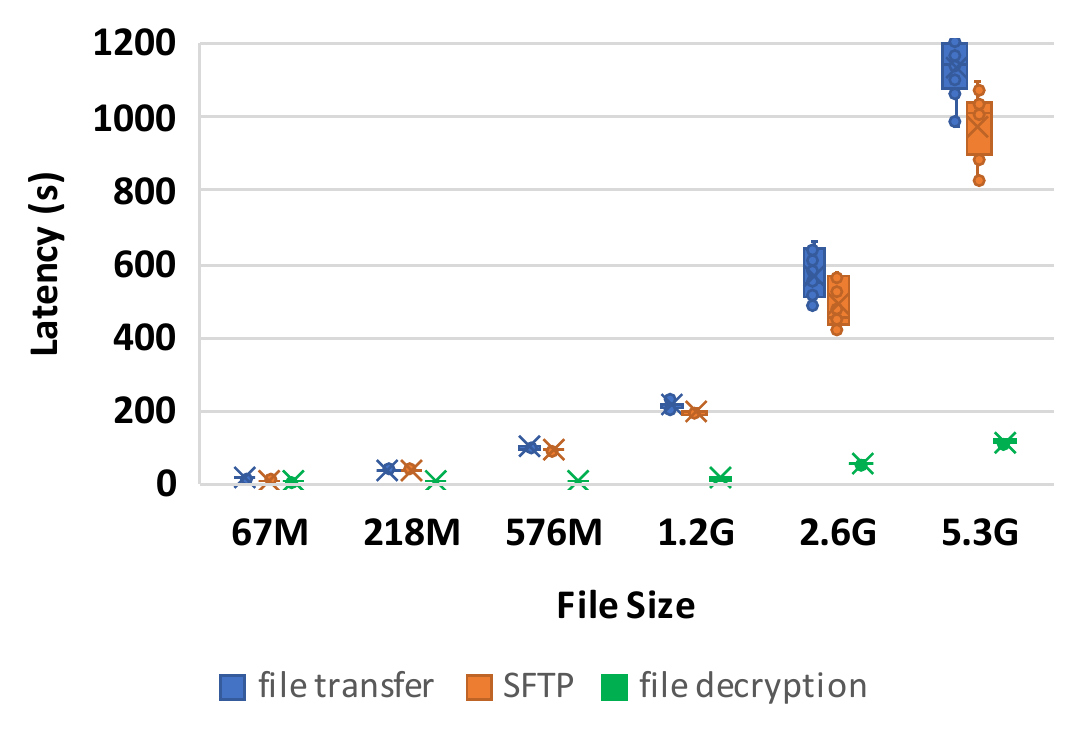}
\caption{SFTP latency and transaction Latency of ``Tx: file transfer" and ``Tx: file decryption"}
\label{fig::1par_tran_dec}
\end{minipage}
\hspace{0.1cm}
\begin{minipage}[c]{0.45\columnwidth}
\centering
\includegraphics[height=0.78\textwidth]{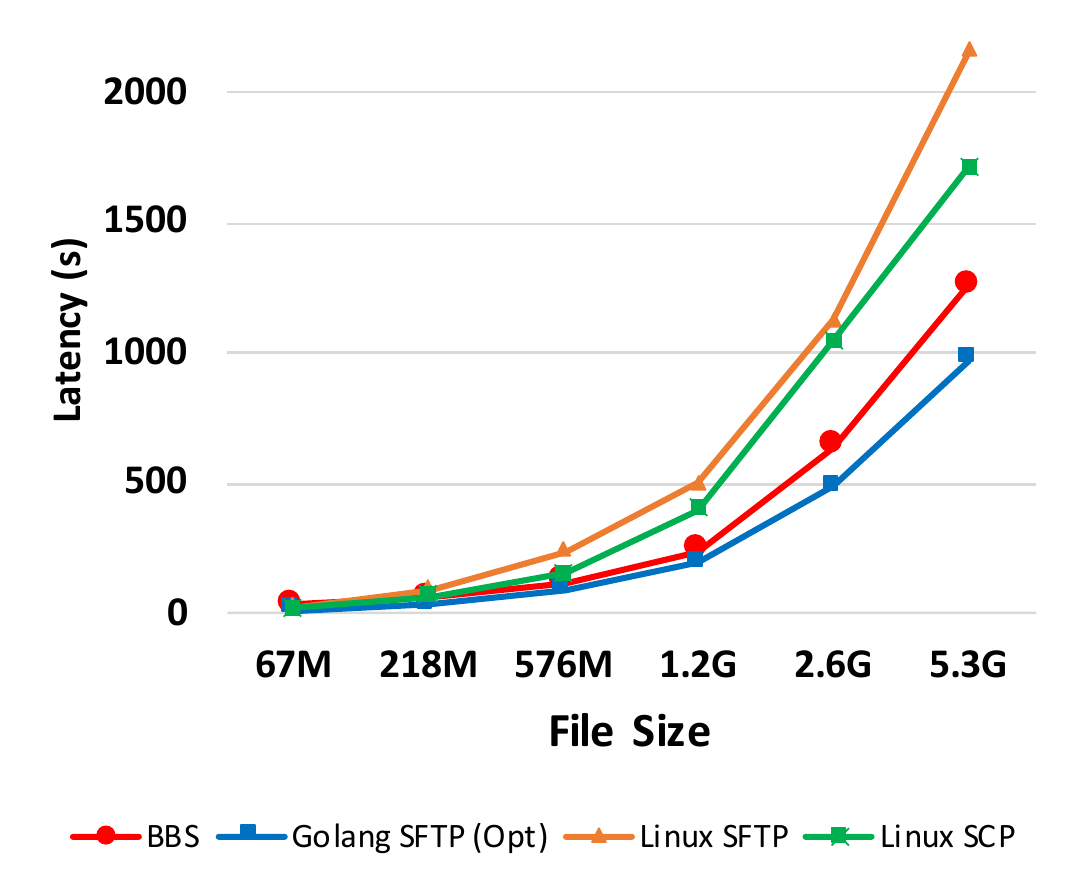}
\caption{File sharing session latency}
\label{fig::1par_whole}
\end{minipage}
\hspace{0.1cm}
\begin{minipage}[c]{0.48\columnwidth}
\centering
\includegraphics[height=0.72\textwidth]{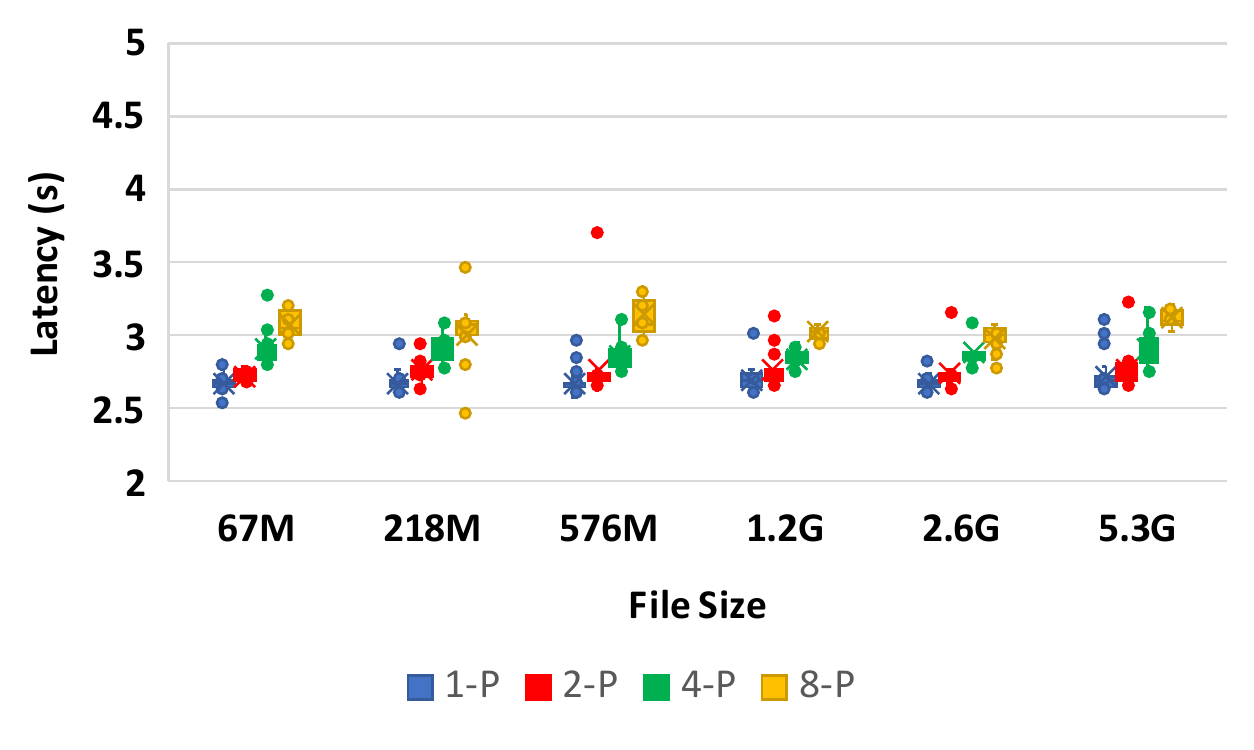}
\caption{Transaction latency of ``Tx: transfer request" in different parallel transfer cases}
\label{fig::mul_par_requestAgree_Tx} 
\end{minipage}
\end{figure*}

\begin{figure*}[!ht]
\begin{minipage}[c]{0.45\columnwidth}
\centering
\includegraphics[height=0.7\textwidth]{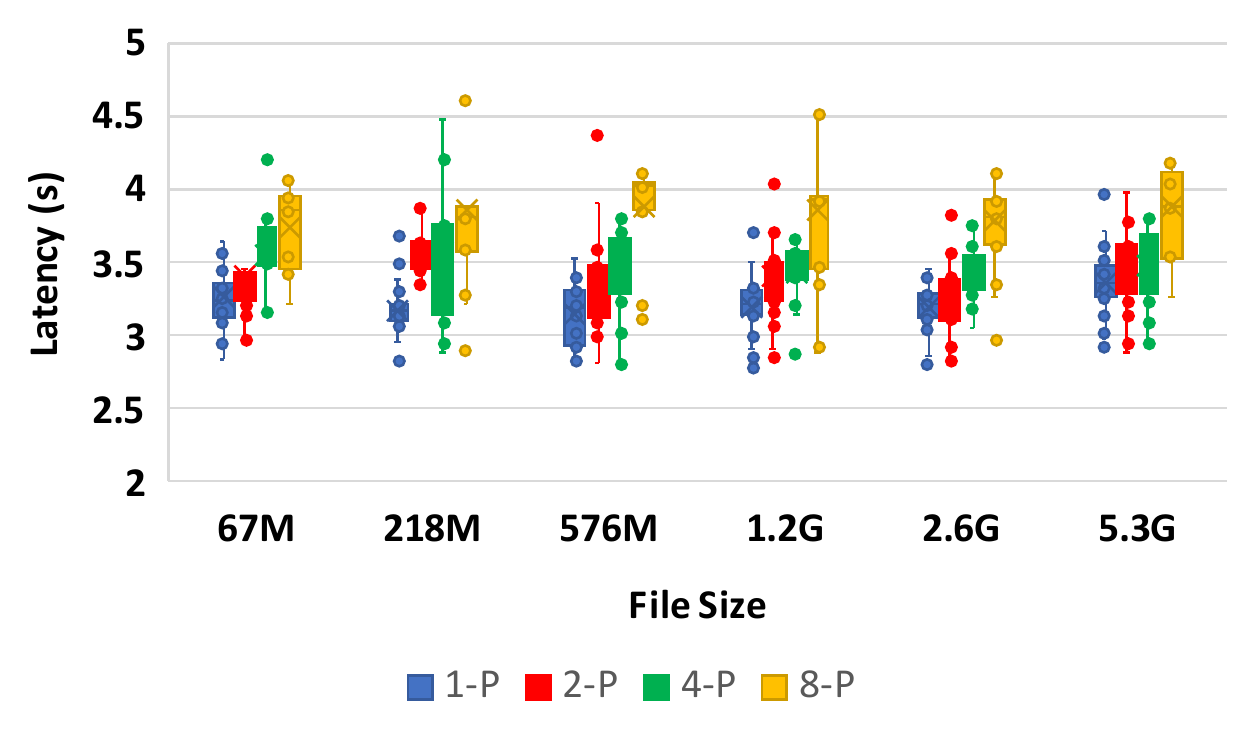}
\caption{Transaction latency of ``Tx: key access" in different parallel transfer cases}
\label{fig::mul_par_requestKey_Tx} 
\end{minipage}
\hspace{0.15cm}
\begin{minipage}[c]{0.45\columnwidth}
\centering
\includegraphics[height=0.7\textwidth]{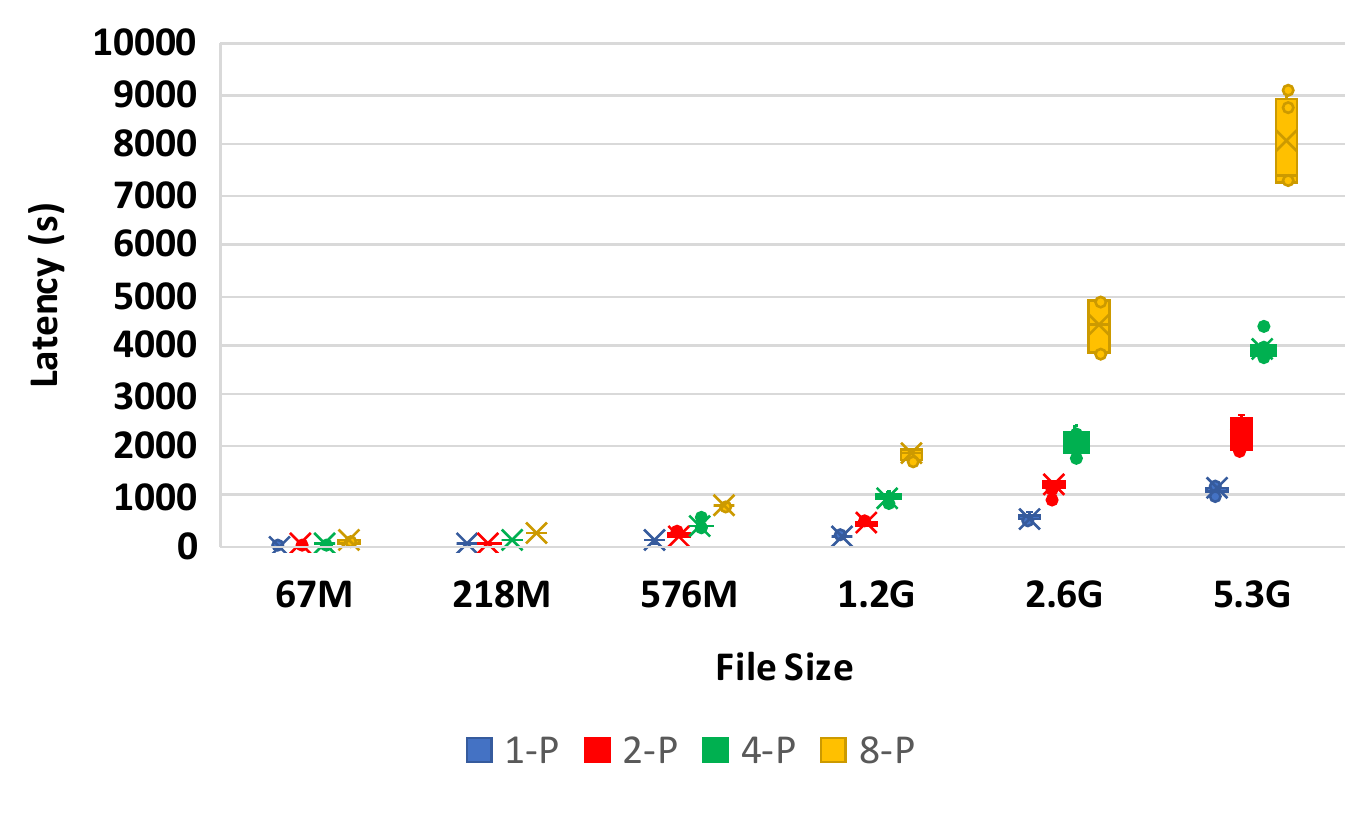}
\caption{Transaction latency of ``Tx: file transfer" in different parallel transfer cases}
\label{fig::mul_par_transfer_Tx}
\end{minipage}
\hspace{0.15cm}
\begin{minipage}[c]{0.45\columnwidth}
\centering
\includegraphics[height=0.7\textwidth]{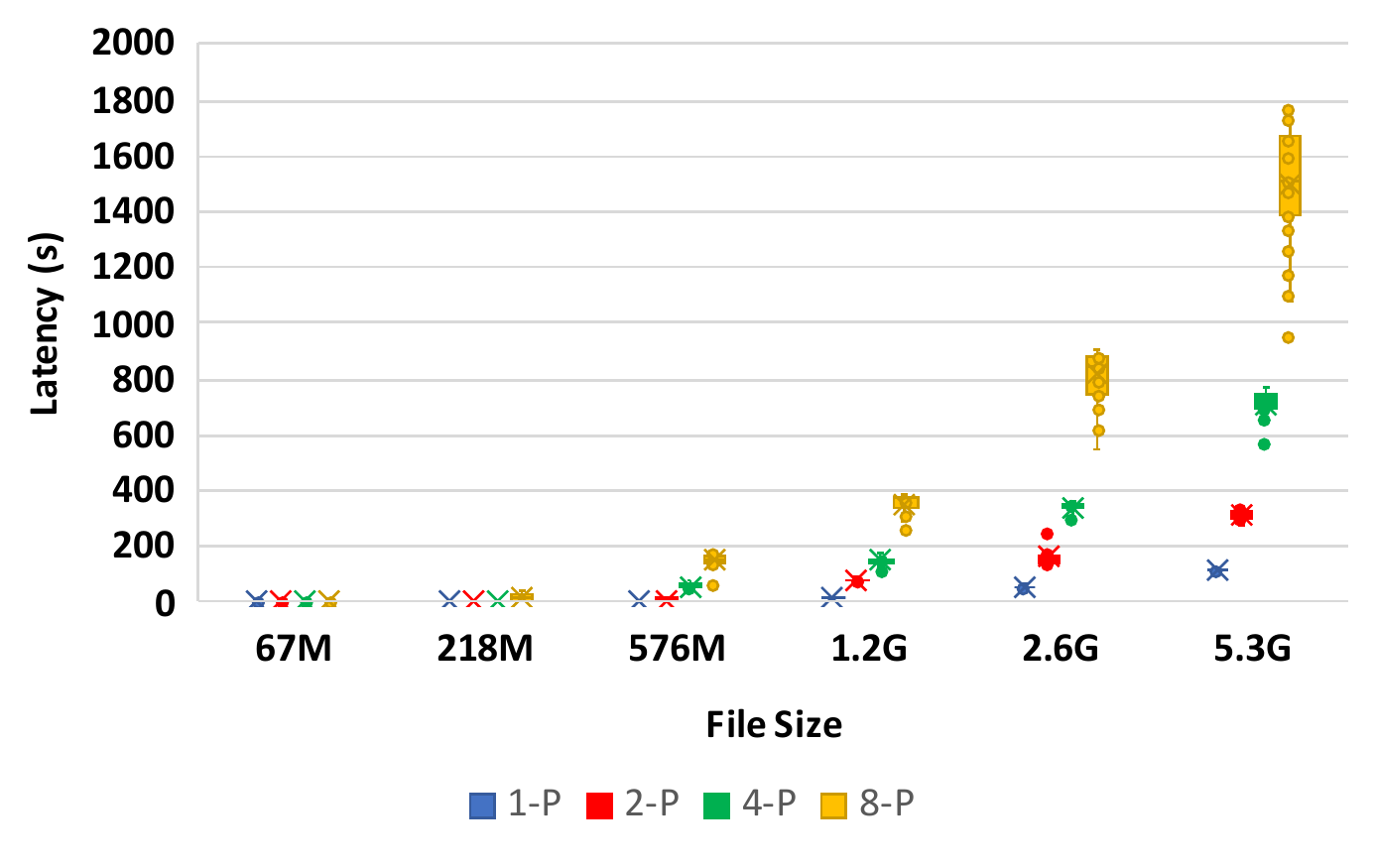}
\caption{Transaction latency of ``Tx: file decryption" in different parallel transfer cases}
\label{fig::mul_par_decrypt_Tx} 
\end{minipage}
\hspace{0.2cm}
\begin{minipage}[c]{0.5\columnwidth}
\centering
\includegraphics[height=0.75\textwidth]{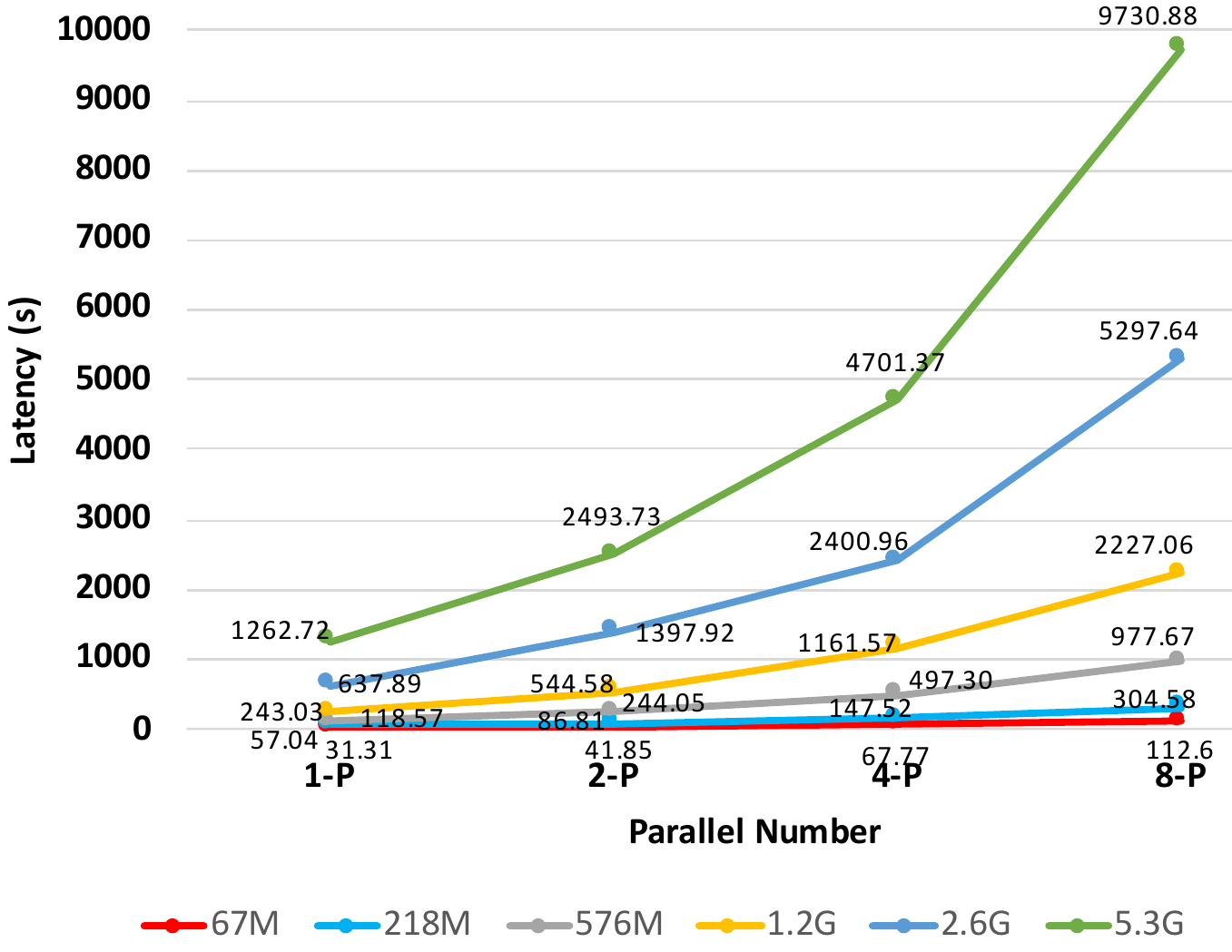}
\caption{Average file sharing session latency in different parallel transfer cases}
\label{fig::mul_par_ave_event} 
\end{minipage}
\end{figure*}


\subsubsection{\textbf{File Size}}
\label{sec::file_size_eva}

For each file in Table \ref{table::file_list}, we conduct the file sharing session 32 times. Each time only one file sharing session is performed. 
The results are shown in Figs. \ref{fig::1par_req}--\ref{fig::1par_whole}. The file size has minor effect on the latency of ``Tx: transfer request" and ``Tx: key access". This is reasonable since these two transactions only operate on the world state data.
The latency of ``Tx: file transfer" and ``Tx: file decryption" rapidly increase as the file size increases.
The reason is that these two transactions involve multiple cryptographic calculations and the Golang \textit{SFTP} operation.
According to Fig. \ref{fig::1par_tran_dec} and Table \ref{table::sftpPercentTran}, the Golang \textit{SFTP} latency accounts for a large proportion of the whole ``Tx: file transfer" latency. As is well know, the \textit{SFTP} latency can be optimized by improving the network bandwidth.

Fig. \ref{fig::1par_whole} shows the latency of one file sharing session of BBS, and compares it with the latency of Linux \textit{SFTP} and \textit{SCP} applications. 
BBS uses the optimized \textit{Golang} \textit{SFTP} with a buffer size of $1$MB. Its latency includes the latency of file transfer via \textit{Golang} \textit{SFTP} and latency incurred by the cryptographic computation and transaction processing and transmission. 
It can be observed that the latency of BBS rapidly increases as the file size increases. This is because normally more time is needed to transfer larger files.
It can also be observed that BBS performs similarly or better compared with Linux SFTP/SCP applications.


\subsubsection{\textbf{Parallel Transfer}}
\label{sec::par_eva}

We also demonstrate that BBS can process multiple big file transfer sessions and other transactions in parallel. 
A long transaction involving the big file transfer does not stop other transactions from running. 
For each file, the number of simultaneous parallel file sharing sessions is 1, 2, 4 and 8, denoted as 1-P, 2-P, 4-P and 8-P in Figs. \ref{fig::mul_par_requestAgree_Tx}--\ref{fig::mul_par_ave_event}, which show the experiment results.
In each case, the file is transferred 32 times in total. For example, in the case of 4-P (4 parallel file sharing sessions), we perform the experiments 8 times and the file is transferred 32 times ($4\times8$).\looseness=-1

We make the following observations. As the number of parallel sessions increases, the latency of ``Tx: transfer request" and ``Tx: key access" is relatively stable as shown in Fig. \ref{fig::mul_par_requestAgree_Tx} and Fig. \ref{fig::mul_par_requestKey_Tx}. However the latency of ``Tx: file transfer" and ``Tx: file decryption" increases as the number of parallel sessions increases. 
Table \ref{table::sftpPercentTran} shows the percentage of \textit{SFTP} incurred latency in the ``Tx: file transfer" latency. The \textit{SFTP} latency accounts for the most of the transaction latency. \looseness=-1

Fig. \ref{fig::mul_par_ave_event} shows the average latency of one file sharing session in different parallel transfer cases. It can be observed that the average file sharing session latency increases as the number of parallel sharing sessions increases. This is reasonable because simultaneous parallel file sharing sessions share the network bandwidth. More sessions mean less network bandwidth for each file sharing session.


%% file: sections/sec9_conclusion.tex
\section{Conclusion}
\label{sec::Conclusion}

In this paper, we solve a novel problem---how to share sensitive big data within a blockchain system autonomously and with no charge, and establish the chain of custody of the shared data securely.
Such a data sharing application is critical for protecting intellectual property (IP) and fighting industrial espionage in fields including biomedical research.
Three challenges including storage space limitation, privacy requirement and security requirement are identified in implementing the blockchain big-data sharing system (BBS).
We denote data such as a big file stored at a blockchain node but outside of the ledger as off-state.
We carefully present our off-state sharing protocol. The transactions generated by our protocol will serve as auditing evidences for the chain of custody.
We implement BBS over Hyperledger Fabric and conduct extensive experiments to evaluate its feasibility and performance.\looseness=-1
